\documentclass[showpacs,aps,pre,twocolumn,amsmath,amssymb,epsfig]{revtex4-1}

\usepackage{graphicx}
\usepackage{epsfig}
\usepackage{dcolumn}
\usepackage{bm}
\newcommand     {\beq}[1]         { \begin{equation} #1 \end{equation} }
\newcommand     {\beqa}[1]        { \begin{eqnarray} #1 \end{eqnarray} }

\begin{document}

\title{Avalanche dynamics in higher dimensional fiber bundle models}

 \author{Zsuzsa Danku$^{1}$}
 \author{G\'eza \'Odor$^{2}$}
  \author{Ferenc Kun$^{1}$}
 \email{Corresponding author: ferenc.kun@science.unideb.hu}
  \affiliation{${}^{1}$Department of Theoretical Physics, University of Debrecen,
 P.O. Box 5, H-4010 Debrecen, Hungary}
 \affiliation{${}^{2}$MTA-MFA-EK,  Center  for  Energy  Research  of  the  Hungarian  Academy  of  Sciences,
H-1121  Budapest,  P.O.\ Box  49,  Hungary}

\begin{abstract}
We investigate how the dimensionality of the embedding space affects 
the microscopic crackling dynamics and the macroscopic response
of heterogeneous materials. Using a fiber bundle model with localized load sharing 
computer simulations
are performed from 1 to 8 dimensions slowly increasing the external load up to failure.
Analyzing the constitutive curve, fracture strength and avalanche statistics of bundles 
we demonstrate that a gradual crossover emerges from the universality class 
of localized behavior to the mean field class of fracture as the embedding dimension 
increases. The evolution between the two universality classes 
is described by an exponential functional form.
Simulations revealed that the average temporal profile of crackling avalanches 
evolves with the dimensionality of the system from a strongly asymmetric shape to 
a symmetric parabola characteristic for localized stresses and homogeneous stress 
fields, respectively.
 \end{abstract}
\maketitle

\section{Introduction}
A large variety of heterogeneous materials respond to a slow external driving 
in a jerky way where sudden outbreaks of activity are separated by silent periods 
\cite{sethna_crackling_2001,alava_statistical_2006}.
From the propagation of imbibition fronts in heterogeneous materials \cite{santucci_imbibition_epl2011}, 
through dislocation bursts of plastically deforming crystals \cite{groma_prl_2014},
and Barkhausen noise in ferromagnets \cite{durin_scaling_2000}, to fracture phenomena 
\cite{laurson_evolution_2013,danku_apl_2015}
and earthquakes \cite{dahmen_nature_2011}, crackling noise has been observed over 
a broad range of length scales.
It was found that crackling noise
is characterized by scaling laws, i.e.\ the statistics of the quantities of single 
bursts is described by power law distributions which may be the fingerprint of an underlying phase 
transition \cite{sethna_crackling_2001,miguel_intermittent_2001,durin_scaling_2000}. 

Recently, it has been pointed out that the average temporal profile is a fundamental 
feature of crackling avalanches. Experimental and theoretical studies have revealed that 
the precise shape of the average profile 
of bursts is sensitive to the details of the physics of the system and it encodes 
valuable information about the underlying intermittent dynamics of pulse generation 
\cite{colaiori_advphys_2008,dahmen_brain_aval_prl_2012,zapperi_signature_2005,
papanikolaou_universality_2011,dahmen_nature_2011}.
For the fracture of heterogeneous materials careful experiments have been performed
where the temporal evolution of individual bursts formed at a propagating 
crack front was determined by direct optical
observation using high speed imaging \cite{PhysRevE.81.046116,PhysRevLett.96.045501,
laurson_evolution_2013}. 
These investigations provided symmetric 
parabolic profiles mainly attributed to long range elastic forces acting along 
the crack front \cite{laurson_evolution_2013}. 
Measurements of magnetic noise induced by the dynamic fracture of steal revealed similar
pulse profiles, however, with a right handed asymmetry \cite{kun_structure_2004,danku_apl_2015}.
The front propagation was modeled 
as the driven motion of an elastic line in a disordered environment of pinning centers.
Varying the range of interaction it was found that the degree of asymmetry depends on the 
range of stress redistribution, i.e.\ profiles evolve with the universality class 
of fracture from a strongly asymmetric shape (localized interaction) to a symmetric parabola 
(long range interaction) \cite{laurson_evolution_2013}. Simulation studies of the dynamics 
of breaking bursts in the 
fiber bundle model underlined the general validity of this behavior 
\cite{danku_PhysRevLett.111.084302}.

In the present paper we take the opposite strategy and address the question 
how the dimensionality of the sample affects the fracture process when the range of 
interaction is kept constant. 
We performed computer simulations in the framework of a fiber bundle model 
with nearest neighbor load sharing after local failure events varying the 
dimensionality of the system from 1 to 8. Both on the macro- and micro-scales
the system exhibits a crossover between the universality classes of localized 
behavior and the mean field class of fracture phenomena. We show that 
this evolution is described by a genuine exponential form.
The temporal profile of breaking avalanches can be well described
by the scaling form suggested in Ref.\ \cite{laurson_evolution_2013} where 
the parameters clearly confirm the crossover between the two universality classes.
Our study shows that the upper critical dimension of the fracture of heterogeneous materials 
is infinite in agreement with a recent theoretical prediction  \cite{hansen_lls_dimension_2015}.
We give numerical evidence that the critical exponents change as an exponential function of the 
dimension.

\section{Local load sharing fiber bundle model in 1 to 8 dimensions}
The fiber bundle model provides an efficient modeling framework
for the fracture of heterogeneous materials \cite{pradhan_failure_2010,hansen2015fiber,
kun_extensions_2006}. 
In spite of its simplicity it captures 
the essential ingredients of fracture phenomena
allowing also for analytical solutions for the most important quantities 
\cite{kloster_burst_1997,hidalgo_avalanche_2009-1}.
The classical fiber bundle model consists 
of $N$ parallel fibers which are organized on a regular lattice. 
In $D=1$ fibers are placed
equidistantly next to each other along a line, while in $D=2$ the fibers are 
assigned to the sites of a square lattice of side length $L$. 
The fibers are assumed to have a perfectly brittle behavior, i.e.\ 
they exhibit a linearly elastic response with a Young modulus $E$ up to breaking 
at a threshold load $\sigma_{th}$. The Young modulus is assumed to be 
constant $E=1$ such that the disorder of the material is solely represented 
by the randomness of the breaking threshold $\sigma_{th}$:
to each fiber a threshold value is assigned $\sigma_{th}^i$, $i=1,\ldots , N$ 
sampled from the probability density $p(\sigma_{th})$. In the present calculations 
we used exponentially distributed breaking thresholds 
\beq{
p(\sigma_{th}) = \frac{1}{\lambda} e^{-\sigma_{th}/\lambda}
\label{eq:expdist}
}
over the range $0\leq\sigma_{th}<+\infty$. 
The scale parameter was fixed to $\lambda = 2$ in all the calculations.

In $D\leq 2$ the bundle is loaded in the direction parallel to the fibers,
which represents the uniaxial loading of a bar shaped specimen. 
Under stress controlled 
loading, when the local load on a fiber reaches its failure strength the fiber breaks
and its load has to be redistributed over the remaining intact fibers. We assume 
localized load sharing (LLS) so that the load of a broken fiber is redistributed 
equally over its intact nearest neighbors in the square lattice 
\cite{hidalgo_fracture_2002,raischel_local_2006,PhysRevE.87.042816}. When the breaking 
fiber is entirely surrounded by intact ones in the square lattice, four fibers share the load, 
however, when the breaking fiber is at the perimeter of a growing broken cluster 
(crack) typically 2-3 fibers receive the excess the load. 
As a consequence, high stress concentration builds up along 
the perimeter of cracks and local stress fluctuations develop.

In higher dimensions $D>2$ the generalization of the model is straightforward 
although it does not have a direct relevance for practical applications: the 
fibers are assigned to sites of cubic lattices and the load on them is represented by 
a scalar variable. 
After failure events the load is redistributed
over the intact nearest neighboring sites along the edges of the lattice.
The emerging stress concentration is controlled by the coordination number $z$ 
of the underlying lattice which depends on the embedding dimension as $z=2D$ in our setup.
In all dimensions periodic boundary conditions are implemented in all lattice directions.

Computer simulations were performed by quasi-statically increasing the external 
load $\sigma$ which is realized by increasing $\sigma$ to provoke the breaking 
of a single fiber. After the
fiber has been removed its load gets redistributed according to the rules discussed above.
The enhanced load on the neighboring fibers may induce further breaking which is followed 
again by a local stress redistribution. As the result of this repeated breaking and stress
redistribution the breaking of a single fiber can give rise to an avalanche of breakings.
The loading process stops when a catastrophic avalanche is triggered in which all remaining
intact fibers break.

In order to keep the problem numerically tractable for the dimensions 
$D=1, 2, 3, 4, 5, 6, 7, 8$ the lattice size was set to $L=4084101, 2021, 159, 45, 21, 13, 9, 7$,
which ensures nearly the same number of fibers in all dimensions. 
To obtain reliable results statistical averaging was done over $K=5000$ simulations.

\section{Macroscopic response}
The mean field limit of FBMs is realized by the equal load sharing (ELS) 
of the load of broken fibers over the intact ones. Under ELS conditions all intact 
fibers receive the same amount of load irrespective of their distance from the 
broken one. It follows that no stress fluctuations can emerge, all fibers keep
the same load during the entire loading process.
\begin{figure}
\begin{center}
\epsfig{file=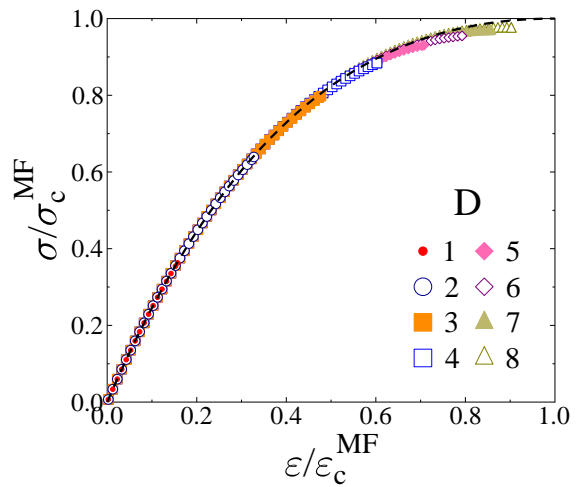,bbllx=30,bblly=20,bburx=380,bbury=330,
width=7.5cm}
\caption{The constitutive curve of the system for different dimensions $D$ scaled with the 
mean field critical strain $\varepsilon_c^{MF}$ and stress $\sigma_c^{MF}$.
The mean field solution Eq.\ (\ref{eq:const_els}) is represented by the dashed line.
}   
\label{fig:constit} 
\end{center}
\end{figure}
Hence, in the mean field limit the random strength of fibers is the only source 
of disorder in the system.
For ELS the macroscopic stress-strain relation $\sigma(\varepsilon)$ 
of the bundle can simply be obtained from 
the general expression $\sigma=E\varepsilon[1-P(E\varepsilon)]$. Here $P(x)$ denotes 
the cumulative distribution of the failure thresholds so that the term $1-P(E\varepsilon)$
yields the fraction of intact fibers which all keep the same load $E\varepsilon$. 
Substituting the exponential distribution Eq.\ (\ref{eq:expdist}) we obtain 
\beq{
\sigma = E\varepsilon e^{-E\varepsilon/\lambda}.
\label{eq:const_els}
}
In Fig.\ \ref{fig:constit} the curve of Eq.\ (\ref{eq:const_els}) is presented 
up to the maximum where catastrophic failure occurs under stress controlled loading.
The constitutive curve has a quadratic maximum the value $\sigma_c$ and position
$\varepsilon_c$ of which determine the fracture stress and strain of the bundle, respectively
\beqa{
\sigma_c^{MF}    &=& \lambda/e, \label{eq:sigc_els}\\
\varepsilon_c^{MF} &=& \lambda/E.
}
Note that the fracture strength $\sigma_c$ and $\varepsilon_c$ of FBMs depend on the 
system size $N$ even in the mean field limit 
\cite{smith_probability_1980,mccartney_statistical_1983,kadar_pre_2017}. 
Although the convergence is rapid, strictly
speaking the above expressions give the bundle strength in the limit of infinite system size.

\begin{figure}
\begin{center}
\epsfig{file=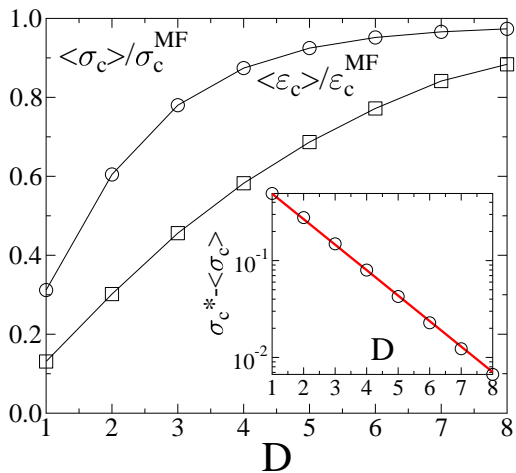,bbllx=10,bblly=10,bburx=350,bbury=300,
width=7.5cm}
\caption{The average fracture stress $\left<\sigma_c\right>$ and strain $\left<\varepsilon_c\right>$
obtained by directly averaging the stress and strain of the last stable configuration 
of the system in stress controlled quasi-static loading simulations. The strength values 
are normalized by their mean field counterpart.
Inset: subtracting a constant from the fracture stress straight line is obtained on a 
semi-logarithmic plot.
}   
\label{fig:last_stable} 
\end{center}
\end{figure}

For finite dimensions $D$ the $\sigma(\varepsilon)$ curves were determined by computer 
simulations averaging 
over a large number of loading processes with different realizations of the threshold 
disorder. It can be seen in Fig.\ \ref{fig:constit} that for all dimensions $D$ the mechanical 
response $\sigma(\varepsilon)$ of LLS bundles follows the mean field solution Eq.\ (\ref{eq:const_els}). 
However, for low dimensional bundles the curves stop significantly earlier implying a 
lower fracture strength and a higher degree of brittleness. As $D$ increases 
the LLS constitutive response completely recovers the mean field behavior.
This tendency becomes more transparent when analyzing the average fracture stress $\left<\sigma_c\right>$
and strain $\left<\varepsilon_c\right>$ as a function of $D$. In
Figure \ref{fig:last_stable} both quantities gradually converge to their mean field 
counterpart, however, the convergence is somewhat faster for $\left<\sigma_c\right>$.
For the highest dimension $D=8$ considered only a few percent difference is observed
from the value of $\sigma_c^{MF}$ of Eq.\ (\ref{eq:sigc_els}).

The most remarkable result is that the convergence to the mean field limit 
can be described by an exponential form. The inset of Fig.\ \ref{fig:last_stable} 
demonstrates that subtracting a constant $\sigma_c^*$ from $\left<\sigma_c\right>$ a straight 
line is obtained on a semi-logarithmic representation which implies the functional form 
\beq{
\left<\sigma_c\right>(D) = \sigma_c^* +A\exp{\left(-D/D^*\right)}.
\label{eq:sigc_conv}
}
Here $D^*$ denotes a characteristic value of the dimension.
Formally, $\sigma_c^*$ is a free parameter in Eq.\ (\ref{eq:sigc_conv}) which 
was tuned to $\sigma_c^*=0.725\pm 0.015$ to obtain the best straight line 
in Fig.\ \ref{fig:last_stable}. Note that this value falls very close to $\sigma_c^{MF}=0.735$.
The characteristic dimension was obtained by fitting $D^*=1.65\pm 0.06$.

The results show that in higher dimensions the role of stress fluctuations is
diminishing in the fracture process and the behavior of the system gradually 
approaches the one of the completely homogeneous stress field 
of the infinite dimensional ELS solution.
It follows that, in spite of the highly localized stress redistribution,
in high $D$ the stochastic breaking process is completely controlled by the 
quenched disorder of the failure strength of fibers.

\section{Avalanche dynamics}
The bundle is loaded in a quasi-static way such that the external load is increased to provoke
solely the breaking of a single fiber. In the simulations the cascade of breaking fibers, emerging 
due to the repeated steps of load redistribution and breaking, is followed 
until it stops while keeping the external load fixed. The size $\Delta$ of
avalanches is characterized by the number of fibers breaking in the avalanche.
These breaking avalanches are analogous to crackling bursts measured in experiments 
with acoustic \cite{garcimar_statistical_1997,PhysRevLett.112.115502} or electromagnetic 
\cite{kun_structure_2004,danku_apl_2015} 
techniques. Under simple 
geometrical conditions such as during the slow propagation of a planar crack direct 
optical observation has also proven very successful 
\cite{maloy_local_2006,tallakstad_local_2011,laurson_evolution_2013}.

\begin{figure}
\begin{center}
\epsfig{file=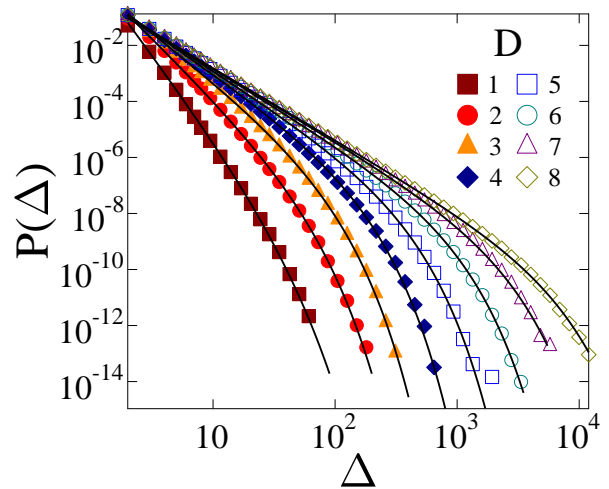,bbllx=10,bblly=10,bburx=350,bbury=300,
width=7.5cm}
\caption{The size distribution of bursts $p(\Delta)$ for all dimensions
considered. The continuous lines represents best fits with the functional 
form Eq.\ (\ref{eq:avalsizedist}).
}
\label{fig:avalsizedist} 
\end{center}
\end{figure}

The probability distribution $p(\Delta)$ of avalanche sizes is presented in Fig.\ 
\ref{fig:avalsizedist}.
In all dimensions $D$ the probability density $p(\Delta)$ is described by the same 
functional form, i.e.\ power law distributions are obtained followed by an exponential 
cutoff
\beq{
p(\Delta) \sim \Delta^{-\tau}\exp{\left(-\Delta/\Delta^*\right)},
\label{eq:avalsizedist}
}
where both the exponent $\tau$ and the characteristic burst size $\Delta^*$
depend on the dimensionality $D$ of the bundle. In Fig.\ 
\ref{fig:avalsizedist} Eq.\ (\ref{eq:avalsizedist}) provides excellent 
fits of the numerical data where the exponent $\tau$ decreases while 
$\Delta^*$ increases with $D$. The result implies that in higher dimensions 
the system can tolerate larger avalanches without catastrophic collapse.
It has been shown by analytical calculations that in mean field FBMs the 
size distribution exponent takes the value $\tau_{MF}=5/2$ which has proven to be 
universal for a broad class of threshold distributions 
\cite{kloster_burst_1997,hansen2015fiber,hidalgo_avalanche_2009}. 
\begin{figure}
\begin{center}
\epsfig{file=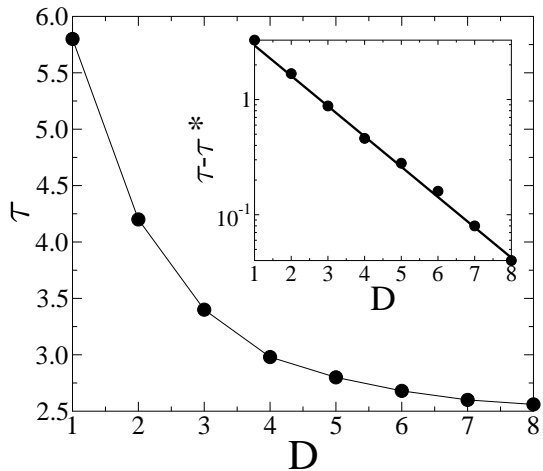,
bbllx=10,bblly=10,bburx=350,bbury=300, width=7.5cm}
\caption{The size distribution exponent $\tau$ of breaking avalanches obtained 
by fitting with the form Eq.\ (\ref{eq:avalsizedist}). Inset: subtracting a proper 
value $\tau^*$ from $\tau$ a straight line is obtained on a semi-logarithmic plot.
}
\label{fig:avalsize_para} 
\end{center}
\end{figure}
For our LLS system it can be observed in Fig.\ \ref{fig:avalsize_para} that $\tau$ has high values 
in low dimensions but with increasing $D$ it approaches the mean field exponent.
The inset of Fig.\ \ref{fig:avalsize_para} demonstrates that the convergence 
is again described by an exponential form similar to Eq.\ (\ref{eq:sigc_conv}) 
\beq{
\tau(D) = \tau^*+B\exp{(-D/D^*)},
}
where best fit was obtained with the same value of the characteristic dimension 
$D^*=1.65\pm 0.06$ as for the fracture strength Eq.\ (\ref{eq:sigc_conv}). 
The value of $\tau^*$ providing the best straight line is $\tau^*=2.52\pm 0.04$ 
which falls very close to the mean field exponent $\tau_{MF}$.

\section{Temporal profile of avalanches}
\begin{figure}
\begin{center}
\epsfig{file=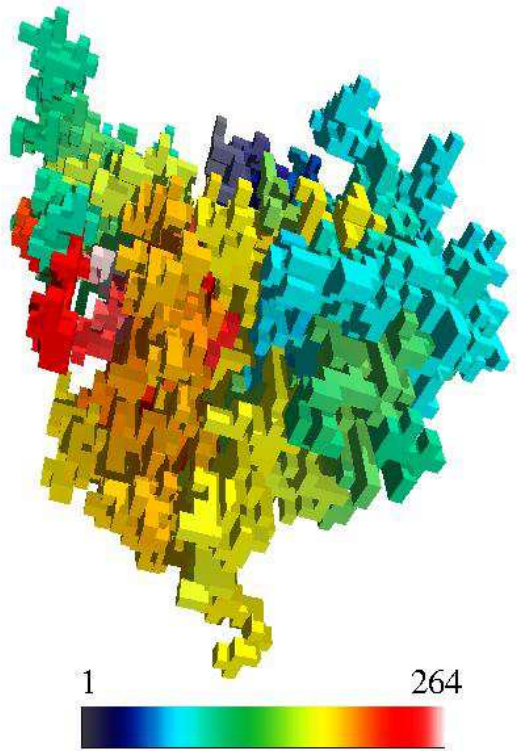,
bbllx=0,bblly=0,bburx=430,bbury=580, width=7.5cm}
\caption{The temporal evolution of a single burst 
of size $\Delta = 8705$ and duration $W=264$ in a three-dimensional bundle.
Single fibers are represented by cubes which are colored according to the 
growth steps they belong to.
}
\label{fig:single_burst} 
\end{center}
\end{figure}

Recently, we have shown that breaking avalanches in fiber bundles have a complex 
time evolution \cite{danku_PhysRevLett.111.084302}: an avalanche typically starts
with the breaking of a single fiber which in turn triggers the breaking of 1-2 additional
fibers after load redistribution. The subsequent load redistribution steps involve a 
larger and larger number of fibers giving rise to a spatial spreading of the avalanche.
The avalanche stops when all the fibers involved in the last redistribution step 
are able to sustain the elevated load. This dynamics implies that avalanches 
are composed of discrete growth steps of size $\Delta_s$, which is the number of fibers 
breaking in a single load redistribution step. The total number of subsequent redistribution -
breaking steps defines the duration $W$ of the avalanche.
The time evolution of a single burst of size $\Delta = 8705$ and duration $W=264$ 
is illustrated in Fig.\ \ref{fig:single_burst} for a three-dimensional bundle, where 
cubes represent fibers. The color code corresponding to the growth steps of the avalanche 
facilitates to follow the breaking sequence.
The temporal evolution of an avalanche is characterized by the $\Delta_s(u)$ function, 
where $u$ is a time variable taking integer values in the interval $u=1,\ldots , W$.
Similarly to the size of bursts $\Delta$, their duration $W$ is also a stochastic 
quantity which varies over a broad range. 
\begin{figure}
\begin{center}
\epsfig{file=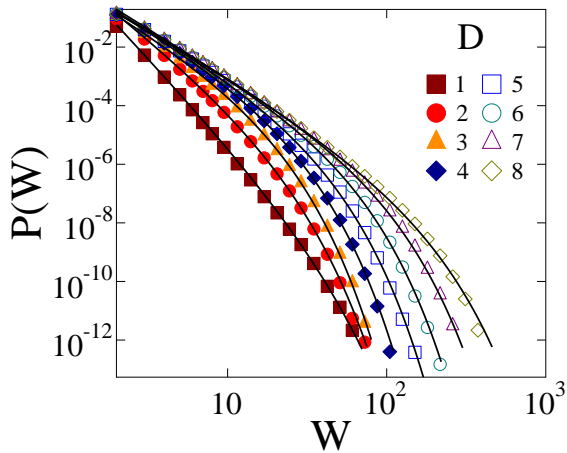,
bbllx=10,bblly=10,bburx=375,bbury=320, width=7.5cm}
\caption{Probability distribution $p(W)$ of the duration $W$ of avalanches 
for different dimensions $D$. The continuous lines represent fits with the 
functional form of Eq.\ (\ref{eq:avalsizedist}).
}
\label{fig:durdist} 
\end{center}
\end{figure}
It can be observed in Fig.\ \ref{fig:durdist} 
that the probability distribution $p(W)$ of the burst duration $W$ has the same 
functional form as $p(\Delta)$, i.e.\ power law behaviour followed by an exponential 
cutoff is evidenced. In higher dimensions
the system can tolerate larger bursts of longer duration, hence, the power 
law exponent $\tau_W$ of $p(W)$ decreases from $\tau_W=5.7$ of $D=1$ to 
the vicinity of $\tau_W\approx 4$ for $D=8$, while the cutoff duration gradually increases
with the dimensionality.

\begin{figure}
\begin{center}
\epsfig{file=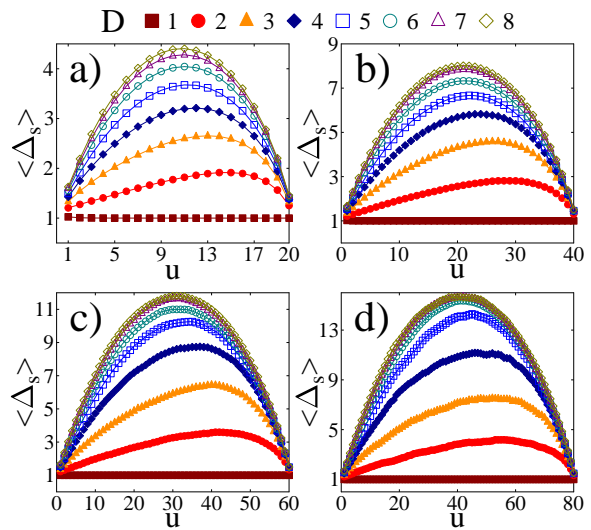, bbllx=10,bblly=10,bburx=720,bbury=640, 
width=7.9cm}
\caption{Average temporal profile of avalanches of different durations for all 
the dimensions $D$ considered ($D$ increases from bottom to top) for four different
durations $W$: 20 $(a)$, 40 $(b)$, 60 $(c)$, 80 $(d)$.
For the special case of $D=1$ the flat profile means that each load redistribution
step causes the breaking of a single fiber. 
For a fixed duration, in low dimensional space the profiles have a strong right handed 
asymmetry which gradually disappears with increasing $D$. 
}
\label{fig:single_burst_profile} 
\end{center}
\end{figure}

For single bursts $\Delta_s(u)$ is a stochastic curve, hence,
quantitative characterization of the internal dynamics of avalanches is provided by the
average temporal profile $\left<\Delta_s(u,W)\right>$, where the step size
$\Delta_s$ is averaged at fixed values of $u$ for avalanches of the same duration W 
\cite{danku_PhysRevLett.111.084302}.
Average profiles are illustrated in Fig.\ \ref{fig:single_burst_profile} 
for different durations $W$ for all dimensions considered. It can be observed
that, except for $D=1$, the shape of avalanches obtained is similar to the experimental findings
\cite{laurson_evolution_2013,danku_apl_2015}. 
For $D=1$ the stress concentration is so high at the tip of growing broken clusters that 
all steps of the breaking sequence have a size $\Delta_s=1$, since a larger number of breaking fibers 
would trigger 
a catastrophic avalanche. Consequently, the emerging pulse profile is completely flat.
At low dimensions $D=2,3$ the profiles have a strong right handed asymmetry at all durations $W$, 
which means that these bursts start slowly, they gradually accelerate while
their stopping is more sudden. 
As the dimensionality of the system increases the degree of asymmetry decreases
and eventually a symmetric parabolic shape is obtained which is characteristic for 
mean field avalanches 
\cite{sethna_crackling_2001,danku_PhysRevLett.111.084302,dahmen_nature_2011}. 
Recently, similar asymmetric avalanche shapes have been obtained in a fiber bundle model 
of creep rupture with localized load sharing \cite{danku_PhysRevLett.111.084302}. Since in those 
calculations the external load was fixed, bursts were triggered by ageing induced 
slow breaking of fibers. However, the cascading breaking sequence of avalanches had essentially
the same dynamics as in the present study.
It has been shown in Ref.\ \cite{danku_PhysRevLett.111.084302} for $D=2$ that the 
short range load sharing and the heterogeneous stress field built up along the
perimeter of cracks are responsible for the right handed asymmetry of pulse profiles.

\begin{figure}
\begin{center}
\epsfig{file=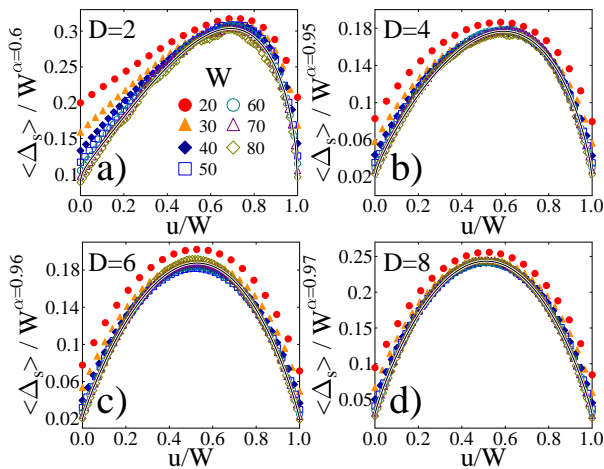, 
bbllx=0,bblly=0,bburx=760,bbury=600, width=7.9cm}
\caption{Rescaling the pulse profiles for a fixed dimension 
with an appropriate power of $W$
profiles of different durations can be collapsed. Except for the shortest duration $W=20$
good quality collapse is obtained which improves with increasing dimension $D$. 
The continuous lines represent fits of the scaling function with Eq.\ (\ref{eq:pulse_profile}).
}
\label{fig:profile_collapse} 
\end{center}
\end{figure}

Comparing the curves in Figs.\ \ref{fig:single_burst_profile}$(a,b,c,d)$ for a fixed dimension 
it can be inferred that bursts of a longer duration $W$ have a larger average height $\left<\Delta_s^{max}\right>$ 
and average size $\left<\Delta\right>$. Figure \ref{fig:profile_collapse} demonstrates that 
rescaling the pulse profiles with an appropriate power of $W$ pulses of different duration 
can be collapsed on the top of each other. The good quality data collapse implies
the scaling form
\beq{
\left<\Delta_s(u,W)\right>=W^{\alpha}f(u/W),
\label{eq:scaling_profile}
}
where the value of the exponent $\alpha$ and the scaling function $f(x)$ both
depend on the dimensionality of the system $D$. In Figure \ref{fig:profile_collapse}
the value of $\alpha$ was tuned to achieve the best collapse.
Deviations from the scaling function $f(x)$ occur for the shortest durations 
which confirms that Eq.\ (\ref{eq:scaling_profile}) is asymptotically valid.

\begin{figure}
\begin{center}
\epsfig{file=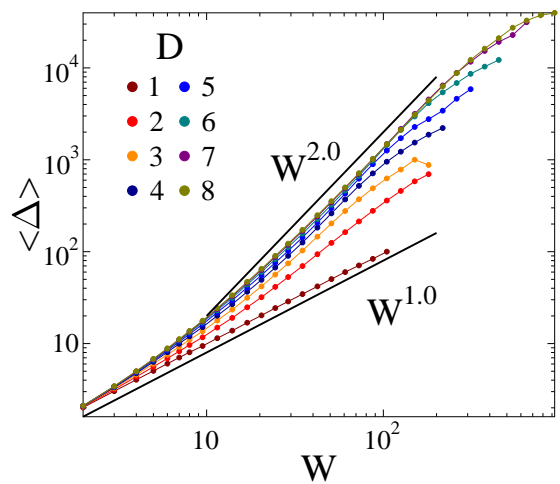, 
bbllx=10,bblly=10,bburx=360,bbury=300, width=7.9cm}
\caption{Average size of bursts as a function of their duration for different embedding
dimensions $D$. The value of the power law exponent increases from 1 to 2 as the dimensionality
increases from 1. The value of $D$ monotonically increases from the bottom curve to the top one.  
}
\label{fig:aver_scale} 
\end{center}
\end{figure}

It follows from the scaling structure Eq.\ (\ref{eq:scaling_profile}) 
that the average avalanche size has a power
law dependence on the duration \cite{danku_PhysRevLett.111.084302,mehta_universal_2002,
sethna_crackling_2001,zapperi_signature_2005,francesca_prl_2003}
\beq{
\left<\Delta\right> \sim W^{1+\alpha}, 
\label{eq:aver_width}
}
which provides an alternative way to determine the exponent $\alpha$, as well.
Figure \ref{fig:aver_scale} shows that the asymptotic behaviour of the average 
size of bursts $\left<\Delta\right>$ of the same duration $W$ can be well described 
by a power law. 
The value of the exponent $1+\alpha$ obtained by fitting 
the $\left<\Delta\right>(W)$ curves with Eq.\ (\ref{eq:aver_width})
is presented in Fig.\ \ref{fig:alpha_tau} as a function of the embedding dimension $D$. 
Since in $D=1$ the step size $\Delta_s$ 
does not exceed 1, the total size $\Delta$ of an avalanche is proportional 
to its duration so that $1+\alpha=1$ and $\alpha=0$ follows. As $D$ increases, $1+\alpha$ 
gradually approaches $2$, and hence $\alpha$ tends to its mean field value $\alpha=1$.
Based on Eq.\ (\ref{eq:aver_width}) a scaling relation 
can be established between $\alpha$ and the exponents $\tau$ and $\tau_W$ of the probability 
distribution of the size and duration of bursts
\beq{
1+\alpha=(\tau_W-1)/(\tau-1) \label{eq:scaling_tau}.
}
In Figure \ref{fig:alpha_tau} 
we compare  the prediction of the scaling law by substituting
the numerical values of $\tau$ and $\tau_W$ on the right hand side of Eq.\ (\ref{eq:scaling_tau}), 
to the value of $1+\alpha$ obtained by direct fitting of the average burst size 
in Fig.\ \ref{fig:aver_scale}.
A reasonable agreement can be observed between the two curves which confirms the consistency
of the results.

\begin{figure}
\begin{center}
\epsfig{file=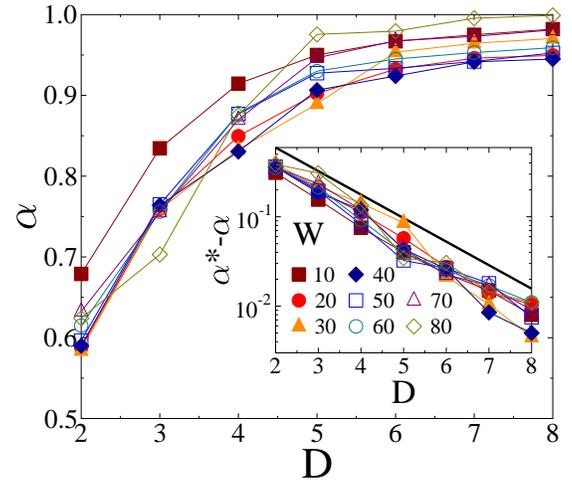, 
bbllx=10,bblly=10,bburx=360,bbury=310, width=7.9cm}
\caption{The $\alpha$ parameter of the avalanche profile described by Eq.\ (\ref{eq:pulse_profile}) 
obtained for avalanches of different durations $W$ as function of the embedding dimension $D$. 
The inset shows that subtracting $\alpha$ from a limit value $\alpha^*$ a straight line is obtained 
on a semi-logarithmic plot which implies an exponential dependence on the dimensionality.
The straight line represents the exponential with the parameter $D^*=1.65$.
}
\label{fig:profile_fit_alpha} 
\end{center}
\end{figure}

\begin{figure}
\begin{center}
\epsfig{file=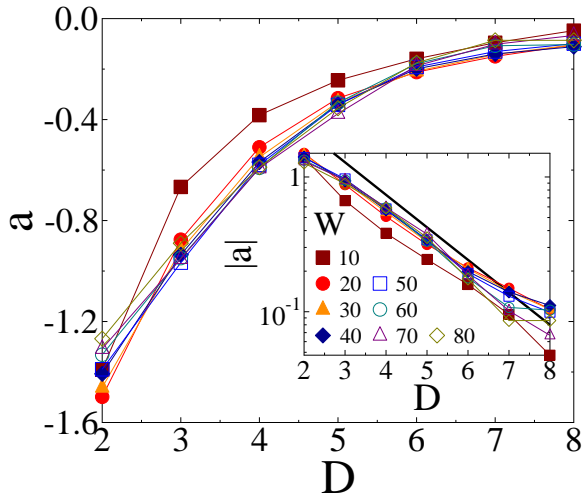, 
bbllx=10,bblly=10,bburx=360,bbury=310, width=7.9cm}
\caption{The asymmetry parameter $a$ of the avalanche profile of Eq.\ (\ref{eq:pulse_profile}) 
obtained for avalanches of different durations $W$ as function of the embedding dimension $D$.
The inset presents the absolute value of $a$ on a semi-logarithmic plot. The straight line implies
that $a$ approaches zero with an exponential dependence on $D$. The straight line represents
the exponential with the parameter $D^*=1.65$. 
}
\label{fig:profile_fit_a} 
\end{center}
\end{figure}

Avalanche profiles $\left<\Delta_s(u,W)\right>$ have an asymmetric functional 
form in all dimensions, however, the degree of asymmetry depends 
on the embedding dimension $D$.
In Ref.\ \cite{laurson_evolution_2013} the following expression has recently been 
suggested to quantify avalanche shapes
\beq{
f(x) \sim \left[x\left(1-x\right)\right]^{\alpha}\left[1-a\left(x-\frac{1}{2}\right)\right].
\label{eq:pulse_profile}
}
Note that the scaling laws Eqs.\ (\ref{eq:scaling_profile},\ref{eq:aver_width})
are consistent with the generic form of Eq.\ (\ref{eq:pulse_profile}) with the same
value of the exponent $\alpha$.
The pulse asymmetry is represented by the parameter $a$
such that zero value of $a$
implies symmetry, while negative and positive values of $a$ capture right and left 
handed asymmetry, respectively. It can be observed in Fig.\ \ref{fig:profile_collapse}
that Eq.\ (\ref{eq:pulse_profile}) provides excellent fits of pulse profiles for all dimensions.
Figures \ref{fig:profile_fit_alpha} and \ref{fig:profile_fit_a} present the value of the 
parameters $\alpha$ and $a$ 
obtained by fitting for several avalanche durations $W$. The careful analysis revealed that
the parameter values practically do not depend on the avalanche duration $W$, except for some
statistical fluctuations the numerical values of $\alpha$ and $a$ agree with each other
at different $W$s. In agreement with the results presented in Fig.\ \ref{fig:alpha_tau}, 
for low dimensions the exponent $\alpha$ starts from the vicinity of $0.6-0.7$
and it increases to $0.95-1$ at high dimensions (see Fig.\ \ref{fig:profile_fit_alpha}). 
At the same time the observed right handed asymmetry of profiles
in Fig.\ \ref{fig:single_burst_profile} gives rise to negative values of $a$ in
Fig.\ \ref{fig:profile_fit_a}. As the event duration $W$ increases from 10 to 80, 
in Figs.\ \ref{fig:profile_fit_alpha} and \ref{fig:profile_fit_a} the numerical error of the determination 
of the parameters $\alpha$ and $a$ increases 
from $0.05$ to $0.15$, and from $0.07$ to $0.16$, respectively. The reason is that 
due to the rapidly decreasing duration distribution $p(W)$ of bursts, the statistics of
events of higher duration decreases.
In higher dimensions $\alpha$ approaches one
and $a$ increases to the vicinity of zero which imply that the profile shape evolves
towards a simple symmetric parabola of mean field crackling systems 
\cite{danku_PhysRevLett.111.084302,dahmen_nature_2011,zapperi_signature_2005,francesca_prl_2003}.
The inset of Fig.\ \ref{fig:profile_fit_alpha} demonstrates that subtracting $\alpha$ from 
a proper limit value $\alpha^*$ straight lines are obtained on a semi-logarithmic plot.
For each $W$ the value of $\alpha^*$ was varied in the range $0.96-1.01$ 
independently until best straight lines were achieved. Similarly, the inset of Fig.\ \ref{fig:profile_fit_a}
presents the absolute value of $a$ where again an exponential dependence is evidenced.
The results imply that the behaviour of the pulse parameters $\alpha$ and $a$ is consistent with 
that of the macroscopic strength and avalanche exponent, i.e.\ they approach their mean field
values with an exponential dependence on the dimensionality. 

\section{Discussion}
The fracture of heterogeneous materials is strongly affected by the degree of disorder,
which controls the spatial variation of microscopic materials' strength, 
and the stress fluctuations emerging due to localized stress redistribution after 
micro-fracturing events. The competition of these two sources of disorders, 
i.e.\ strength and stress disorders,
gives rise to a highly complex fracture process which manifests
itself in the variation of macroscopic strength, in the
statistics of crackling bursts, and in the temporal evolution of single crackling events.
At a fixed amount of quenched strength disorder stress fluctuations are mediated
by the range of load redistribution. Former studies have revealed two universality classes
of fracture, i.e.\ the localized load sharing (LLS) class and the mean field class (ELS),
characterized
by a high degree of brittleness and a quasi-brittle response with a large amount of 
avalanche precursors of failure, respectively.
\begin{figure}
\begin{center}
\epsfig{file=atl_Delta_W_illeszt_param.eps, 
bbllx=50,bblly=35,bburx=375,bbury=327, width=7.9cm}
\caption{The value of the exponent $1+\alpha$ of Eq.\ (\ref{eq:aver_width}) compared 
to the outcome of the scaling relation Eq.\ (\ref{eq:scaling_tau}). The good agreement
of the two curves confirms the consistency of the results.
}
\label{fig:alpha_tau} 
\end{center}
\end{figure}
In order to understand the competing role of different disorder sources between the 
limiting cases of the LLS and ELS classes, in our study we considered a fiber bundle 
model with a fixed amount of strength disorder and varied the dimensionality of the system
from 1 to 8 at a fixed range of load sharing. The strength disorder is represented by an exponential
distribution of the failure threshold of fibers, which falls in the well understood universality
classes both in the ELS and LLS limits of load sharing. In all dimensions nearest neighbor
load redistribution was implemented on cubic lattices with periodic boundary conditions 
in all lattice directions. 

Our study revealed a very interesting dimensional crossover between the two universality 
classes of fracture: Both on the macro and micro-levels fracture characteristics 
evolve with the dimensionality from the highly brittle response of low dimensional systems
controlled by stress fluctuations, to the quasi-brittle behavior in high dimensions where the
strength disorder dominates. For the macroscopic strength of the bundle and for the 
power law exponent of the size distribution of crackling bursts the convergence to the mean 
field limit is described by an exponential functional form.

Avalanche profiles have been found before to be very sensitive to models' details, especially
to the degree of correlations of microscopic events leading to collective avalanches 
of local failures. We showed that the symmetry of avalanche profiles depends on the dimensionality
of the system, gradually shifting from a strongly asymmetric shape at low dimensions to a symmetric 
parabolic form in the mean field limit. The parameters of pulse profiles were found to evolve towards
their mean field values with an exponential dependence on the dimensionality similarly to
the macroscopic strength and avalanche exponent of the system. 

Increasing the dimensionality of the bundle implies a decreasing stress concentration
along broken clusters by increasing the connectivity of the system. 
As a consequence, stress fluctuations
have a diminishing role with increasing dimension giving more room 
for the quenched disorder of fibers' strength. 
The evolution of avalanche temporal profiles 
we observe with the embedding dimension is similar to what has been obtained recently for 
avalanches of a propagating crack front when increasing the range of interaction
in a fracture model of fixed $D=2$ dimensions \cite{laurson_evolution_2013}.

Recently, the effect of the long range connection 
of fibers on the fracture process has been studied. Instead of a regular lattice,
fibers were assigned to the nodes of a complex network with small world 
properties \cite{d.-h.kim_universality_2005}. 
Redistributing the load along the edges of the network simulations revealed 
that a small amount of long range connection is sufficient to converge the system 
from the LLS to the mean field universality class. Eventually, the dominance of 
quenched structural disorder is responsible  for the ELS behavior similarly to our case.

Fiber bundles in higher dimensions have recently been investigated in Ref.\ 
\cite{hansen_lls_dimension_2015}. Focusing on the amount of damage accumulated 
up to failure and on the distribution of avalanches a crossover from the LLS 
to the mean field universality class was pointed out. However, the crossover 
is described by a power law functional form contrary to our exponential 
behavior. This difference may arise due to the different protocols of 
load redistribution used in the simulations. 
In agreement with Ref.\ \cite{hansen_lls_dimension_2015}, our   
results imply that the upper critical dimension of the fracture of heterogeneous materials 
is infinite. Further support of this remarkable feature of fracture phenomena 
could be obtained by a finite size scaling analysis in higher dimensions, however,
it turned to be infeasible due to the overwhelming numerical costs.

In the broader context of critical phenomena, the absence of a finite upper critical
dimension has also been found in Kardar-Parisi-Zhang type surface growth models 
\cite{odor_2008_universality_book},
where the critical exponents present an approximate
exponential dependence on the dimensionality of the system \cite{odor_kpz_pre_2010}.
We conjecture that the quenched disorder of the system and the locally conserving nature 
of the dynamics (i.e.\ the entire load dropped by broken fibers is redistributed over the intact ones 
without loss) are responsible for the absence of a finite upper critical dimension beyond 
which mean field behavior is attained.
The exponential crossover from the local to the mean field universality class of fracture involves 
a characteristic dimension $D^*$ which falls between 1 and 2. Since avalanche shapes of the one 
dimensional system do not conform with the higher dimensional ones, we propose the interpretation
of $D^*$ as the lower critical dimension of fracture phenomena. Our simulations confirm
that for $D>D^*$ all characteristic quantities of the system evolve through gradual quantitative
changes but the qualitative behaviour remains robust as the embedding dimension increases.

\begin{acknowledgments}
The work is supported by the EFOP-3.6.1-16-2016-00022 project. 
The project is co-financed by the European Union and the European Social Fund.
This research was supported by the National Research, Development and
Innovation Fund of Hungary, financed under the K-16 funding scheme Project no.\ K 119967
and K 109577. This work was supported through the New National
Excellence Program of the Ministry of Human Capacities
of Hungary.
The research was financed by the Higher Education Institutional
Excellence Programme of the Ministry of Human Capacities in Hungary, 
within the framework of the Energetics thematic
programme of the University of Debrecen.
\end{acknowledgments}

\bibliography{/home/feri/papers/statphys_fracture}

\end{document}